\documentstyle[12pt]{article}
\def\Section#1{\setcounter{equation}{0}\section{#1}}

\def\la{\mathrel{\mathpalette\fun <}}
\def\ga{\mathrel{\mathpalette\fun >}}
\def\fun#1#2{\lower3.6pt\vbox{\baselineskip0pt\lineskip.9pt
\ialign{$\mathsurround=0pt#1\hfil##\hfil$\crcr#2\crcr\sim\crcr}}}

\newcommand{\bc}{\begin{center}}
\newcommand{\ec}{\end{center}}
\newcommand{\bd}{\begin{displaymath}}
\newcommand{\ed}{\end{displaymath}}
\newcommand{\ba}{\begin{array}}
\newcommand{\ea}{\end{array}}
\newcommand{\bt}{\begin{tabular}}
\newcommand{\et}{\end{tabular}}
\newcommand{\un}{\underline}

\begin{document}

\def\deh{{\pa}_{0}^{\!\!\!\!\leftrightarrow}\,}
\def\disc{\mbox{\,disc}\,}
\def\intf{\int d^{4}x\,}
\def\intgr{\int_{4}^{\infty}}
\def\ints{\int_{4}^{\infty} ds'}
\def\lar{\longrightarrow}
\def\pa{\partial}
\def\pal{p_{\al}}
\def\pbe{p_{\be}}
\def\pga{p_{\ga}}
\def\pde{p_{\de}}
\def\qba{\overline{q}}
\def\roo{(\frac{s-4}{4})}
\def\suz{\mbox{SU(2)}}
\def\sud{\mbox{SU(3)}}
\def\tp{\tilde p}
\def\tr{\mbox{tr}\,}
\def\Tr{\mbox{Tr}\,}
\def\ue{\mbox{U(1)}}
\def\uka{\underline{k}}
\def\upe{\underline{p}}
\def\al{\alpha}
\def\be{\beta}
\def\ga{\gamma}
\def\de{\delta}
\def\ka{\kappa\,}
\def\la{\lambda}
\def\ep{\varepsilon}
\def\om{\omega}
\def\Ph{{\it\Phi}}
\def\psiba{\overline{\psi}}
\def\si{\sigma}
\def\th{\theta}
\def\va{\varphi}

\def\beq{\begin{equation}}
\def\eeq{\end{equation}}
\def\bed{\begin{displaymath}}
\def\eed{\end{displaymath}}
\def\beqq{\begin{eqnarray}}
\def\eeqq{\end{eqnarray}}
\def\bedd{\begin{eqnarray*}}
\def\eedd{\end{eqnarray*}}
\def\Ph{{\it\Phi}}

\def\header{\begin{flushleft}
            ZU-TH 25/94\\October 1995
            \end{flushleft}}
\def\eq{eq.~}

\thispagestyle{empty}
\header \vspace*{2cm}
\bc
\Large\bf
Unitary Background Gauges and Hamiltonian approach\\
to Yang--Mills Theories \footnote{partially supported by
Schweizerischer Nationalfonds.}
\ec

\vspace*{2.0cm}

\centerline{\large A. Dubin$^a$ and D. Wyler$^b$}

\vspace*{0.5cm} \centerline{\large $^a$ Institute of Theoretical
and Experimental Physics, 117259 Moscow, Russia}
\centerline{e-mail address dubin@vxitep.itep.su} \vspace*{0.5cm}
\centerline{\large $^b$ Universit\"at Z\"urich, 8057
  Z\"urich, Switzerland}
\centerline{e-mail address wyler@physik.unizh.ch}

\vspace*{3cm}

\begin{abstract}
A variety of unitary gauges for perturbation theory in a background
field is considered in order to find those most suitable for
a Hamiltonian treatment of the system. We select
two convenient gauges and derive the propagators $D_{\mu\nu}$
for gluonic quantum fluctuations immersed in background configurations.
The first one is a unitary generalization of the usual
Coulomb gauge in QED which preserves the decoupling of
two propagating polarizations from the instantaneous one. The second
possibility is the axial light cone gauge which remains
ghost free also in the presence of a background. Applications of the
formalism to the spectrum and dynamics of QCD at the confinement scale,
such as hybrid states, are  briefly discussed.
\end{abstract}

\newpage

\Section{Introduction}

The identification of relevant degrees of freedom \cite{1,2,3}
in the non-perturbative region of QCD is the crucial step
towards a consistent theory of hadrons.
One of the ways \cite{1,2} to disentangle this problem in an
economical way is to  separate in Euclidean space first non-trivial 'vacuum'
configurations $B_\mu$, e.g. to split the total gluonic field $A_\mu$
into a background and a quantum fluctuation

\beq
\label{gl1}
A_\mu = B_\mu +a_\mu
\eeq
where the field $B_{\mu}$ is primarily responsible for the long
range gluonic correlation functions.
The qualitative picture for $B_{\mu}$--fields is the ensemble of lumps
described by their collective coordinates. For the
self consistency of the separation (\ref{gl1}), these
collective coordinates should
not be distorted after inclusion of quantum fluctuations $a_{\mu}$.

Recently, this idea has been emphasized \cite{4,5}, with the suggestion
to consider perturbation theory in a confining background which leads
to the area law asymptotics of the averaged Wilson loop.  The (path)
integral over the background fields can be reformulated \cite{14} as a
summation over irreducible correlators of vacuum field strength
tensors by
means of the cluster expansion. In this way, the existence of flux tubes
between quarks is related to decaying correlators with
particular Lorentz structures \cite{14}. Although the procedure appears
reasonable,
and is supported by the results from the simpler $2+1$
compact $U(1)$ theory \cite{2} where the monopole background fields
generate at large distances the 'frozen' string,
no consistent theory exists for such background fields in QCD.
At present, one must {\em postulate} that the
dynamics of the vacuum configurations
leads to a finite string tension, which can be expressed as the
infinite sum over contributions from the vacuum correlators of all
orders.

Since the confining background cures all infrared
singularities \cite{4}, it enables one to investigate the dynamics
of the $a_\mu$--perturbations at large distances as well.
The important point is that
the $a_\mu$--fluctuations immersed into
such vacuum fields describe the effective
long distance excitations
of the string \cite{4,5} frozen at the level of the vacuum contribution.
They should be considered, besides quarks, as the relevant
degrees of freedom at the scale of
confinement. For instance, the interpretation of the
one  $a_\mu$-gluon exchange amplitude (averaged over the
$B_\mu$-fields) of Fig.1a requires introducing the first
excitation of the string as in Fig. 1b. In general, the
QCD string between quarks will be divided by valence
gluons into a corresponding number of elementary
'frozen' pieces.

Within this picture, the basic question is how to describe
the gluonic excitations ('constituent gluons') $a_\mu$
interacting via the background in a  nonperturbative way.
Because of gauge invariance,
$a_\mu$ in eq.(\ref{gl1})
contains (unphysical) pure gauge components.
Also the presence of non-decoupled propagating ghosts inherent in
the standard formulation \cite{6,1,2} of perturbation theory in
a background field  with the usual gauge
condition

\beq
\label{gl2}
D_\mu(B) a^\mu = 0
\eeq
will obscure
the analysis of bound states. It is the main goal of this
paper to explore these problems and to select
unitary background gauges where
ghosts are either absent or nonpropagating.  This will allow us to
describe bound states in terms of physical polarizations of $a_{\mu}$ only.

To apply the propagators
 $D_{\mu\nu} (B) = <a_{\mu}a_{\nu}>$ to a given
problem, we suggest to use the multichannel
Hamiltonian approach \cite{7,8} recently generalized \cite{9}
to this kind of
fluctuating strings.  It provides a scheme to derive from the
propagator both the diagonal elements of the Hamiltonian
for states with a fixed
number of propagating constituent gluons as well as the nondiagonal
terms which mix different Fock states. In this way, we can
investigate glueballs or hybrids.
We stress that $D_{\mu\nu}(B)$
provides us with both
spin-dependent and spin-independent nonperturbative interactions
between $a_{\mu^{-}}$gluons and other constituents at the scale of
confinement. We note that the diagonal approximation for such
states based on the standard nonunitary background gauge (\ref{gl2}) (with
the propagating ghosts) has recently been suggested in ref. \cite{5}.

The plan of the paper is as follows. In the next section,
we discuss further the special role of the unitary background
gauges for the analysis of $QCD$ bound states. Then, in section 3, we
review the derivations of $D_{\mu\nu}$ in the
abelian case within the path integral
approach.

In section 4, the standard analysis of  perturbation
theory in background fields is extended to prove the well known
background gauge invariance \cite{6,1,2} for the case of the special
unitary gauges to be described in sections 5 and 6. There,  the
explicit form of the propagators $D_{\mu\nu}(B)$ will be derived.
Finally section 7 contains a brief sketch of the applications and
the conclusions.

\Section{The special role of unitary background gauges}

Before constructing explicitly the propagators in the unitary background
gauges, we  discuss in more detail why they are important for the analysis of
gluonic excitations in bound states.

As stressed above, the key point is
that if dynamical gluons appear as
constituents of a bound state, such as in glueballs or
in hybrid states, non-physical polarizations of $D_{\mu\nu}$ (
which occur in a
nonunitary gauge) obscure both the interpretation of the quantum numbers
of a hadron {\em and} the dynamical scheme of interactions between the
valence constituents.  Due to gauge invariance, the contributions
from 'pure gauge' polarizations should drop out (together with ghost
contributions) in the final result.  But in order to invoke
physical intuition, it is important to have the transparent and
economical
formulation which is provided by
unitary gauges where ghosts are either absent or
nonpropagating. Then all independent propagating polarizations in
$D_{\mu\nu}$ can be interpreted as physical degrees of freedom.

This special role is illustrated already at the level of standard
perturbation theory if a bound state is involved.
In the calculation of deep inelastic scattering amplitudes,
it is the axial or planar gauge (see \cite{10} for the refs.)
in which only the planar
Feynman graphs contribute to the leading logarithmic asymptotics
which allows for a selfconsistent introduction of the hadronic
structure functions. In such a gauge, there is  a direct correspondence
between properly chosen independent fields and the "physical" degrees
of freedom. This conceptual
necessity of a particular gauge stands in contrast to the usual
perturbative analysis where the expansion is restricted to a given
order in the coupling constant which makes results explicitly gauge
invariant. In that case, the choice of gauge is essentially a
technical matter.

Another example of a dynamical scheme where  gluons become
'constituents' is the approaches based on the light-cone $QCD$-Hamiltonian
\cite{8}. As in any quantum field theory, a multichannel Hamiltonian
must be introduced in order to reproduce, via iterations, the
amplitudes which initially are expanded in time ordered Feynman
graphs. In the absence of ghosts, wave functions of Fock
states with extra dynamical gluons will describe only the "least"
number of physical polarizations. In this
approach \cite{8}, the unitary light cone gauge
(for the standard perturbation theory)
insures this property.


To apply effectively the multichannel Hamiltonian approach to the
dynamics of gluonic excitations in a background we will
therefore look for
gauges where  \\
i) ghosts are either absent or  do not propagate, \\
ii) there is an unambiguous and simple way to separate
in $D_{\mu\nu}(B)$ the non-propagating polarization
from the propagating ones.

A gauge satisfying condition i) will be called unitary;
it reproduces correctly the number of independent physical
propagating polarizations interacting with the background fields. The
second requirement gives an additional selection of unitary gauges
leading, as we will
see, to the most economical use of the Hamiltonian approach.

Gauge invariance obviously allows to eliminate one of the four
components of the $a_{\mu}$-gluons in a background.
But since they are generally not
'on mass-shell', it is by no means evident how to formulate a
generalized transversality condition by which only two propagating
polarizations are retained and which is equivalent to the
non-propagation of ghosts.
We will show in this work that there are two most suitable unitary
gauges and derive the corresponding propagators $D_{\mu\nu}(B)$.

The first one generalizes the ordinary Coulomb gauge propagator of $QED$
and is based on the condition

\beq
\label{gl3}
N_i(B) a_i\equiv(D_i(B)+2D^{-1}_0(B)\hat F_{0i}(B)) a_i=0.
\eeq
leading to a propagator $D_{\mu\nu}(B)$ which satisfies
\beq
\label{gl4}
D_{0i}(B) =0,\qquad i=1,2,3 .
\eeq
Here $D^{ab}_{\mu}(B)$ and $\hat
F^{ab}_{\mu\nu} \equiv f^{abc}F^c_{\mu\nu}(B)$ are the standard covariant
derivative and field strength tensor respectively.

The $D_{00}(B)$ part corresponds to the instantaneous Coulomb
exchange modified
by the interaction with the background. As for the spatial
components $D_{ik}(B)$, we will show that
the generalized projector

\beq
\label{gl5}
P^{ab}_{ik}= \delta_{ik}\delta^{ab}-(N_i(B)N^{-2}_l(B)N_k(B))^{ab}
\eeq
allows one to represent $D_{ik}(B)$ in the sandwiched form
\beq
\label{gl6}
D_{ik}(B) = P_{im}(B) (K(B)^{-1})_{mn} P_{nk}(B).
\eeq
This gives  the correct transversality
condition for the propagating physical components orthogonal  to the
instantaneous polarization in accordance with eq. (\ref{gl4}).

All other unitary modifications of the Coulomb gauge in the presence of a
background do not admit the decoupling condition (\ref{gl4}) important for the
further applications of $D_{\mu\nu}$. As a result, requirement ii)
is not satisfied.

The second unitary gauge can be directly obtained from the
light cone gauge constraint

\beq
\label{gl7}
(n_\mu a_\mu)^a =0 ,\qquad n_\mu^2=0.
\eeq
Other axial gauges (with $n_\mu^2 \neq 0$) being unitary in the
presence of background fields do not satisfy condition
ii).

The transversality condition of eq. (\ref{gl6}) for the
nonperturbatively interacting
gluons has two immediate consequences. First it
gives selection rules for the allowed quantum numbers of states
with a fixed number of gluonic constituents. A well known
example is the (Landau--Yang) theorem for the forbidden
total momentum of the system of two real photons.
Similarly, this condition
decreases the dimensions
of multiplets with fixed allowed quantum numbers.

\Section{Path integral derivation of QED propagator \newline
in Coulomb and
axial gauges}

Let us first review the procedure of gauge fixing performed
directly in the path integral representation for the generating
functional $Z(J)$.

We will formulate the criterion for a gauge to be ghost free in a
way which can be easily generalized to the perturbation theory in a
nonabelian background. Our starting point is the standard
expression in Euclidean space

\begin{equation}
\label{gl8}
Z(J) = \int DA_{\mu}|_{G.F.} \exp \left[- \int d^4x \left(
\frac{1}{4g^2} F^2_{\mu \nu} + J_{\mu} A_{\mu} \right) \right]
\end{equation}
and we impose the gauge invariance condition

\begin{equation}
\label{gl9}
\partial_{\mu} J_{\mu} = 0.
\end{equation}
To evaluate $Z(J)$ in the form

\begin{equation}
\label{gl10}
Z(J) \sim \exp \left[ -\frac{g^2}{2} \int d^4x d^4y
J_{\mu} (x) D_{\mu \nu} (x-y) J_{\nu}(y) \right]
\end{equation}
one must separate explicitly in eq.(\ref{gl8}) the gauge zero modes
$A^{(0)}_{\mu}$ \begin{equation}
\label{gl11}
A^{(0)}_{\mu} = \partial_{\mu} g
\end{equation}
of the
quadratic term $F^2_{\mu\nu}$
or, in other words, fix a gauge. The standard way is to
insert into eq. (\protect\ref{gl8}) the
Faddeev--Popov unity in the form

\begin{equation}
\label{gl12}
1 = det\frac{\delta f(A^{\omega}_{\mu})}{\delta \omega} \int Dg \delta
(f(A^g_{\mu}))
\end{equation}
where $A^g_{\mu} = A_{\mu} + \partial_{\mu}g$ and $f(A_\mu)$ is the gauge
fixing function satisfying $f(A^{\omega}_\mu)=0$. Thus
\begin{equation}
\label{gl13}
Z(J) = \int DA_\mu det
\left( \frac{\delta f(A^{\omega}_\mu)}{\delta \omega}\right)
\delta (f(A_\mu)) \exp \left[ - \int d^4 x \left(\frac{1}{4g^2}
F^2_{\mu \nu} + J_{\mu} A_{\mu} \right) \right],
\end{equation}
where we omitted the volume of gauge modes $\int Dg$.

By fixing the gauge in this way, the number of independent polarizations
in $D_{\mu\nu}$ is always decreased from four to three. There remains the
freedom to select one non-propagating polarization.

Rewriting $F^2_{\mu\nu}$ via the kinetic matrix $K_{\mu\nu}$ as
\bd
2A_{\mu}K_{\mu\nu}A_{\nu}
\ed
and taking for simplicity  zero sources $J_{\mu}=0$, one gets

\beq
\label{gl14}
Z(0)=\left[\frac{det\frac{\delta f}{\delta\omega}}{det
K_{ik}}\right]^{1/2}.
\eeq
Here, $i$, $k$ refer to the three components of $K_{\mu\nu}$ selected
by $\delta(f(A))$.

Instead of counting independent propagating
polarizations of $D_{\mu\nu}$, the absence of ghosts can
conveniently be formulated as follows:
$det K_{ik}$ should have a factorized form
\begin{eqnarray}
\label{gl15}
det K_{ik}&=& det \left( \frac{ \delta f}{ \delta\omega} \right)
\widetilde{det} \left(P_{im} \tilde K_{mn}P_{nk} \right) \nonumber \\
\frac{ \partial}{ \partial( \partial^2_{ \mu})}
\frac{ \delta f}{ \delta \omega} &=& 0 \mbox{}
\end{eqnarray}
where the operator
$(\frac{\delta f}{\delta\omega})^{-1}$ must be non-propagating
and the corresponding determinant in eq.(\ref{gl14}) must
cancel exactly the ghost factor in the numerator of eq. (\ref{gl14}).
We use the notation $\widetilde{det}$ in order to stress that
the last determinant in eq. (\ref{gl15})
denotes formally the result of the integration over
the two propagating physical
polarizations (selected by the projector
$P_{ik}$ ($P_{im}P_{mk}=P_{ik}$)). They determine
completely (if eqs.(\ref{gl15}) are satisfied)
the statistical sum $Z(0)$

\beq
\label{gl16}
Z(0)=\left[\widetilde{det}\left(P_{im}\tilde K_{mn}P_{nk}\right)\right]^{-1/2}.
\eeq

In QED, the ghosts decouple always and the first condition of eq.
(\ref {gl15}) holds; only the second one is nontrivial. In
the presence of a nonabelian background $B_\mu$,
the ghosts interact with the
$B_{\mu}$-fields and factorization (\ref {gl15}) of the
determinant in the statistical sum  $Z(0,B)$ is not
insured for every gauge.
This is related to the fact that in  gaussian approximation
$Z^{(2)}(0,B)$ of
the statistical sum  is not invariant under the choice of background gauges for
quantum fluctuations. But a background gauge is nevertheless
unitary (ghosts do not propagate) if the second condition of
eq. (\ref{gl15}) is met.

In the remainder of this section our aim is to gain experience in how to
use the freedom in gauge fixing for bringing the propagator $D_{\mu
\nu}(x-y)$ into the form the most convenient for the Hamiltonian
technique.

\subsection*{A) Coulomb gauge }

Instead of imposing from the begining the gauge
condition (\ref{gl12}) in the form

\beq
\label{gl17}
f(A_i)=\partial_iA_i
\eeq
one can start from the 'physical' conditions and
require  that $D_{\mu \nu}$ satisfies

\begin{equation}
\label{gl18}
\label{condi}
 \frac{\partial}{\partial(\partial^2_0)} D^{col}_{00}(\partial_{\mu}) = 0,
\quad D^{col}_{0i}(\partial_{\mu}) = D^{col}_{i0}(\partial_{\mu}) = 0,
\quad i = 1, 2, 3
\end{equation}
which we would like to preserve in the presence of a background.
The second condition of eqs.(\ref{gl18}) implies that the spacial and
temporal components, $A_i$ and
$A_0$, are decoupled,  while the first one insures that  $A_0$
does not propagate in time (and is responsible for the instantaneous
interaction).

Since  there are only two transverse
polarizations for ordinary photons, one can impose an
additional constraint on
$D_{ik}(\partial_{\mu})$ via insertion of the gauge unity (\ref{gl12})
in a form which leads to eqs.(\ref{gl18}).
To understand which function $f(A_{\mu})$ brings $D_{\mu\nu}$ into
the required form, we first write
$\int d^4 x \left(\frac{1}{4} F^2_{\mu \nu}
\right)$ as

\begin{equation}
\label{gl19}
\int d^4 x \frac{1}{4} F^2_{\mu \nu} = \frac{1}{2} \int d^4
x (-A_0 \partial^2_i A_0 + 2A_0 \partial_0 \partial_i A_i -
\{A_i \partial^2_{\mu} \delta_{ik} A_k - A_i \partial_i \partial_k
A_k \}).
\end{equation}
Comparing with (\ref{gl18}), we see that the
mixed term $A_0 \partial_0 \partial_i A_i$ should be canceled.
This immediately leads to the standard  form (\ref{gl17}) of $f(A_{\mu})$.

Summarizing, we see that starting from conditions (\ref{gl18})
one can selfconsistently
determine the form of gauge fixing function $f(A)$ leading to the
ghost free gauge.  In Section 4 we will generalize this idea to
the case of perturbation theory in a nonabelian background.

The main steps leading to $Z(J)$ which will
be repeated in the nonabelian case are as follows.
In order to get rid of $\delta(\partial_i A_i)$, we introduce
the projector
on transverse states

\begin{equation}
\label{gl20}
P_{ik} = \left( \delta_{ik} - \frac{\partial_i
\partial_k}{\partial^2_l} \right), \quad \partial_i P_{ik} = 0.
\end{equation}
Decomposing into transverse and longitudinal parts
\beq
\label{gl21}
A_i=P_{ik}A_k+(1-P)_{ik}A_k \equiv A^{\perp}_i+A^{\parallel}_i,
\eeq
one obtains

\begin{equation}
\label{gl22}
\delta(\partial_i A_i) = (det (-\partial^2_i))^{-1/2} \delta ((1 -
P)_{ik} A_k).
\end{equation}
Therefore, the integration over $(1-P)_{ik} A_k \equiv A^{\parallel}_i$ is
eliminated and  $Z(J)$ takes the form

\begin{eqnarray}
\label{gl23}
& & Z(J)= \int DA_0 DA^{\perp}_i (det(-\partial^2_i))^{1/2}
\exp \left[+\frac{1}{2} \int d^4 x \left( \frac{1}{g^2}
(A_0 \partial^2_i A_0 + \right. \right. \nonumber \\
& & + \left. \left. A^{\perp}_i \partial^2_{\mu} A^{\perp}_i) - 2(J_0 A_0 +
J^{\perp}_i A^{\perp}_i) \right) \right].
\end{eqnarray}
Because only the transverse components $J^{\perp}_i$ enter,
the resulting propagator
will have  the sandwiched form
\begin{equation}
\label{gl24}
D_{ik} = P_{il} \tilde{D}_{lm} P_{mk},
\end{equation}
given after integration over $A_0$ and $A^{\perp}_i$
by  the final representation (\ref{gl10}) for $Z(J)$

\begin{equation}
\label{gl25}
Z(J) = (det (-\partial^2_{\mu}))^{-1} \exp \left[ -\frac{g^2}{2}
\int d^4x \left(J_i P_{il}\frac{1}{\partial^2_{\mu}} P_{lk} J_k +
J_0 \frac{1}{\partial^2_i} J_0   \right) \right].
\end{equation}
We stress that the remnant of the ghost determinant in eq.(\ref{gl23})
has been cancelled by
the integration over the instantaneous component $A_0$.
The expression for the statistical sum (3.9)
\begin{equation}
\label{gl26}
Z(0) = (det (-\partial^2_{\mu}))^{-1}
\end{equation}
is to be interpreted as that for two (transverse) propagating
polarizations.

\subsection*{B) Axial gauge with $n^2\ne0$}

In the abelian case (or in the case of standard nonabelian
perturbation expansion) there is another conventional ghost free
gauge in which the propagator includes only two
propagating polarizations,
the well known axial gauge $(nA) = 0$. The gauge fixing
function can be conveniently  choosen  as

\begin{equation}
\label{gl27}
f(A) = (n\partial)(nA).
\end{equation}
so that the second condition in eq. (\ref{gl15}),
written in terms of light cone variables, holds.
Here, $n$ is an arbitrary 4-vector. If it is light-like, the
corresponding gauge is the
light cone gauge.

All $n^2\ne0$ are conceptually equivalent to the elementary cases
$n=(1,\vec0)$ and $n=(0,1,\vec{0}_{\perp})$. The derivation
of the propagator is similar in
both; we discuss only

\begin{equation}
\label{gl28}
n_{\mu} = (1, \vec{0}).
\end{equation}
In full analogy with the previous subsection, one obtains
\begin{equation}
\label{gl29}
 Z(J)= \int DA_i (\det(-\partial^2_0))^{1/2}
\exp \left[+\frac{1}{2} \int d^4 x
( \frac{1}{g^2} A_i K_{ik} A_k - 2J_i A_i) \right]
\end{equation}
where
\begin{equation}
\label{gl30}
K_{ik} = \partial^2_{\mu} \cdot P_{ik} + \partial^2_0 (1-P)_{ik}.
\end{equation}
and the projector $P_{ik}$ is defined by eq. (\ref{gl20}). The
representation (\ref{gl30}) for
$K_{ik}$ provides the
separation of nonpropagating and propagating polarizations such
that integration over the former cancels the ghost factor
$det^{1/2}(-\partial^2_0)$. This separation is
analogous to that of eq. (\ref{gl23});
$A^{||}_i$ represents the nonpropagating component.

Rewriting $DA_i$ as $DA^{\perp}_i DA^{\parallel}$, we arrive at the
final representation
\begin{equation}
\label{gl31}
Z(J) = (det (-\partial^2_{\mu}))^{-1} \exp \left[ -\frac{g^2}{2}
\int d^4x \left( J_i (P \frac{1}{\partial^2_{\mu}} P)_{ik} J_k +
J_i ((1 - P) \frac{1}{\partial^2_0} (1 - P))_{ik} J_k \right) \right].
\end{equation}

In accordance with gauge invariance, $Z(0)$ has the same form (and
interpretation) as in the Coulomb gauge.
We note that in $D_{\mu\nu}$ the nonpropagating part
represented by the last term
in the exponent of eq.(\ref{gl31}) replaces
the instantaneous Coulomb exchange
part of eq. (\ref{gl25}).

\subsection*{C) Light cone gauge, $n^2=0$}

To complete the analysis, we consider also the light cone axial gauge
$(nA) = 0$

\begin{equation}
\label{gl32}
n_- = 1, \quad n_+ = n_{\perp} = 0
\end{equation}
where the light cone coordinates for a 4-vector $b_{\mu}$ are
 $b_{\pm} = \frac{1}{\sqrt{2}}(b_z \pm b_0)$ and  we
choose the gauge fixing function $f$ as before.

This choice of $n$ requires a Minkowski metric and straightforward
calculations lead to the following analogue of eqs.(\ref{gl23}) and
(\ref{gl29})
\begin{equation}
\label{gl33}
Z(J) = \int DA_{-} DA_{\perp} (det (-\partial_{+}^2))^{1/2}
\exp \left[+\frac{i}{2}\int d^4x\left(\frac{1}{g^2}A_{\mu}K_{\mu\nu}
A_{\nu} + 2 J_{\mu} A_{\nu} \right) \right]
\end{equation}
where $\mu, \nu$ run over $+,-, \perp = \{i,k\}$ and
\begin{equation}
\label{gl34}
K_{++} = - \partial^2_{+}, \; K_{+i}= K_{i+} = - \partial_{+}
\partial_i, \; K_{ik} =  (2 \partial_{+}\partial_{-} +
\partial^2_{\perp})\delta_{ik} - \partial_i \partial_k.
\end{equation}
The form (\ref{gl34}) of $K_{++}$ implies that  the
integration over the non-propagating component
$A_-$ exactly cancels the ghost factor
$(det(-\partial^2_+))^{1/2}$ with the result
\begin{eqnarray}
\label{gl35}
& & Z(J) = \int DA_{\perp} \exp \left[ +\frac{i}{2g^2} \int d^4x
\left( [A_i
K_{ik}
A_k + 2g^2J_i A_i] \right. \right.  + \nonumber \\
& & + \left. \left. \{g^2(Jn) -(A_{\perp}\partial_{\perp} )(n\partial)\}
\frac{1}{(n \partial)^2} \{g^2(Jn) - (n \partial) (\partial_{\perp}
A_{\perp}) \} \right) \right] .
\end{eqnarray}
Consequently, the final form of $Z(J)$ reads

\begin{eqnarray}
\label{gl36}
& & Z(J) = (det (-\partial^2_{\mu}))^{-1} \exp \left[ +
\frac{ig^2}{2} \int d^4 x \left[(Jn) \frac{1}{(n \partial)^2}
(Jn) \right. \right. -  \nonumber \\
& & - \left. \left. \left( J_i - (Jn) \partial_i
\frac{1}{(n \partial)} \right)  \frac{1}{\partial^2_{\mu}}
\left( J_i - \frac{1}{(\partial n)} \partial_i (Jn) \right) \right] \right]
\end{eqnarray}
which corresponds to
\label{gl37}
\begin{equation}
D_{\mu \nu} = \left[ -g_{\mu \nu} + \frac{\partial_{\mu} n_{\nu} +
\partial_{\nu} n_{\mu}}{(n \partial)} \right]
\frac{1}{\partial^2_{\mu} + i \varepsilon}.
\end{equation}
The sum in the exponent of eq.(\ref{gl36}) naturally provides us with the
separation of propagator (3.30) into the instantaneous
\begin{equation}
\label{gl38}
D^{inst}_{\mu \nu} =+ \frac{n_{\mu} n_{\nu}}{(n \partial)^2}
\end{equation}
and propagating parts
\begin{equation}
\label{gl39}
D^{prop}_{\mu \nu} = \left[ -g^{\perp}_{\mu\nu} +
\frac{n_{\mu} \partial^{\perp}_{\nu} +
n_{\nu}\partial^{\perp}_{\mu}}{(n\partial)} -
\frac{n_{\mu} n_{\nu} \partial^2_{\perp}}{(n\partial)^2} \right]
\frac{1}{\partial^2_{\mu} + i \varepsilon}.
\end{equation}
This  will be useful when we extend our considerations to the
nonabelian background.
It is easy to check that
 the transversality condition is satisfied
\begin{equation}
\label{gl40}
N_{\mu} D^{prop}_{\mu\nu} = 0, \quad N_{\mu} = \partial^{\perp}_{\mu}
+ m_{\mu} (n \partial)
\end{equation}
with the new 4-vector $m_{\mu}$
\begin{equation}
\label{gl41}
m_{+} = 1, \quad m_{-} = m_{\perp} = 0
\end{equation}
which implies that there are only two propagating
polarizations as before.

Summarizing, we see that eliminating in $Z(J)$ one
polarization by inserting the gauge unity (\ref{gl12}), one
generates in all three
cases a corresponding ghost determinant which  is exactly
canceled by the integration over the retained
physical 'nonpropagating'
polarization. In other
words, the  determinant
\begin{equation}
det (K_{\mu\nu}) \nonumber
\end{equation}
of the kinetic matrix for the three retained polarizations
assumes a factorized form (\ref{gl15}) where the
part which cancels the ghost determinant can be  separated
from the one describing
two propagating physical polarizations.

\Section{Gauge fixing procedure for the perturbation expansion in a
nonabelian background}

As in the abelian case, we begin with the path integral form
\cite{6,11} of the generating functional in the presence of a
background $B_{\mu}$

\begin{equation}
\label{gl43}
Z(J,B) = \int Da_{\mu} \det \frac{\delta f(B^{\omega},
a^{\omega})} {\delta \omega}|_ {\omega_0} \delta (f(B,a))
  \exp \left[ - \int d^4 x \left(
\frac{1}{4g^2} (F^a_{\mu \nu} (B+a))^2 + J^a_{\mu} a^a_{\mu} \right)
\right],
\end{equation}
where $ f(B^{\omega}, a^{\omega})|_{\omega_0}=0$
keeping in mind unitary gauges.
In what follows the standard notations \cite{11} are used:
\begin{equation}
\label{gl44}
F_{\mu \nu} (B+a) \equiv F_{\mu\nu}^a T^a = F_{\mu\nu} (B) +
D_{\mu}(B) a_{\nu} - D_{\nu}(B) a_{\mu} - i [a_{\mu}, a_{\nu}]
\end{equation}
with

\begin{equation}
\label{gl45}
D_{\mu}^{ab} = \partial_{\mu} \delta^{ab} + f^{abc}B^c_{\mu} =
\partial_{\mu} \delta^{ab} - i(T^c)^{ab} B^c_{\mu} \equiv
\partial_{\mu} \delta^{ab} - i(B_{\mu})^{ab}
\end{equation}
\begin{equation}
\label{gl46}
[a_{\mu}, a_{\nu}] \equiv a^a_{\mu} a^b_{\nu} [T^a, T^b] =
i f^{abc} T^c a^a_{\mu} a^b_{\nu}.
\end{equation}
For zero sources $J_{\mu} = 0$, the action density
$\sim F^2_{\mu\nu}(B+a)$ obeys the gauge symmetry
\beq
\label{gl47}
(B_{\mu} + a_{\mu}) \rightarrow U(B_{\mu} + a_{\mu} +
i \partial_{\mu}) U^{-1}
\eeq
which in the infinitesimal form reads

\begin{equation}
\label{gl48}
B_{\mu} + a_{\mu} \to B_{\mu} + a_{\mu}+ D_{\mu}(B+a)\omega,
\end{equation}
and where $D_{\mu}(A)\omega$ can be expressed according to eq.(4.3) as

\begin{equation}
\label{gl49}
T^a(\partial_{\mu} \omega^a + f^{abc} \omega^b A^c) \equiv
T^a (\partial_{\mu} \omega^a + (\omega \times A)^a).
\end{equation}

There are two related issues which distinguish the
gauge fixing procedure in eq.(\ref{gl43}) from the one for the standard
perturbation theory with no background.

The first one is the well known fact \cite{6} that in the
presence of $B_{\mu}$ it is possible to select  for
the $a_{\mu}$-fields a gauge which
preserves the invariance of $Z(J,B)$ under gauge transformations of
$B_{\mu}$ and the sources $J_{\mu}$

\begin{equation}
\label{gl50}
B_{\mu} \rightarrow U (B_{\mu} + i\partial_{\mu}) U^{-1}, \quad
J_{\mu} \rightarrow UJ_{\mu} U^{-1}.
\end{equation}

Usually this invariance is exploited within the background field
method \cite{6,11} to simplify the renormalization scheme and
multiloop calculations. Here we would like to stress  the
integration over background configurations resulting \cite{14}
in particular
properties of the asymptotics for averaged Wilson loops.  To build up
the formalism in terms of gauge invariant objects like Wilson loops
(with possible spin-insertions, see for example \cite{12}), it is
necessary to
represent physical amplitudes in terms of
gauge invariant combinations of Greens
functions in the background fields such as $<\bar{\Psi}(x) \Psi(y)>$
and $<a_{\mu}(x)a_{\mu}(y)>$ for the the matter and the $a_{\mu}$-fields,
respectively.
Consequently, it is economical to work with such a gauge for
$a_{\mu}$ in which the Greens function  $D_{\mu\nu}(x,y|B)=<a_{\mu}(x) a_{\nu}
(y)>$  is  transformed homogeneously with respect to background gauge
transformation (\ref{gl50})

\beq
\label{gl51}
D_{\mu\nu} (x,y|B) \to U(x) D_{\mu\nu}(x,y|B) U^{-1}(y).
\eeq

This property follows if $Z(J,B)$ is invariant under (\ref{gl50}).

The second aspect is  the freedom
to split the gauge  variations of
$B_{\mu} + a_{\mu}$ field  in eight
ways between $B_{\mu}$ and $a_{\mu}$ fields:

\begin{equation}
\label{gl52}
\delta^{(1)} a_{\mu} = D_{\mu}(B+a)\omega, \quad
\delta^{(1)} B_{\mu}= 0
\end{equation}

\begin{equation}
\label{gl53}
\delta^{(2)} a_{\mu} = D_{\mu}(B)\omega, \quad
\delta^{(2)} B_{\mu} = (\omega \times a_{\mu})
\end{equation}

\begin{equation}
\label{gl54}
\delta^{(3)} a_{\mu} = (\omega \times a_{\mu}), \quad
\delta^{(3)} B_{\mu} = D_{\mu}(B)\omega
\end{equation}

\begin{equation}
\label{gl55}
\delta^{(4)} a_{\mu} = D_{\mu}(a) \omega, \quad
\delta^{(4)} B_{\mu} = (\omega \times B_{\mu})
\end{equation}

\begin{equation}
\label{gl56}
\delta^{(5)} a_{\mu} = (\omega \times B_{\mu}), \quad
\delta^{(5)} B_{\mu} = D_{\mu}(a) \omega
\end{equation}

\begin{equation}
\label{gl57}
\delta^{(6)} a_{\mu} = \partial_{\mu} \omega, \quad
\delta^{(6)} B_{\mu} = (\omega \times (B_{\mu}+a_{\mu}))
\end{equation}

\begin{equation}
\label{gl58}
\delta^{(7)} a_{\mu} = (\omega \times (B_{\mu}+a_{\mu})), \quad
\delta^{(7)} B_{\mu} = \partial_{\mu} \omega
\end{equation}
where we omit the eighth irrelevant splitting
$\delta^{(8)} a_{\mu} = 0$.

Since no sufficiently comprehensive discussion of these
points exists in the
literature \cite{6,11}, we will present the detailed analysis required
to construct the nonstandard unitary gauges in the next two
sections.

First, we single out the
splittings leading to a $Z(J,B)$ which is invariant under (\ref{gl50}).
Since the  action of eq.(\ref{gl43}) itself is invariant under all
seven forms of the variation (the third one, (4.12), corresponding to that of
eq. (\ref{gl50})),  possible non invariances must come from the
gauge fixing determinant $\det \frac{\delta^{(i)} f(B^{\omega},
a^{\omega})}{\delta \omega}$ of the Faddeev-Popov unity in eq. (\ref{gl43}).
Thus, in order to insure invariance under
(\ref{gl50}), $\frac{\delta^{(i)} f}{\delta \omega}$ must
transform homogeneously under
the  infinitesimal  variation  (\ref{gl54}) (with
infinitesimal parameter $\tilde{\omega}$), i.e.

\begin{equation}
\label{gl59}
\delta^{(3)} \left( \frac{\delta^{(i)} f(B^{\omega},a^{\omega})}
{\delta \omega} \right)
= \left( \tilde{\omega} \times \frac{\delta^{(i)}
f(B^{\omega},a^{\omega})}{\delta \omega}
\right).
\end{equation}
For simplicity, we first take covariant gauges and consider
the function $f(B,a)$ (linear in $a_{\mu}$) written in terms
tranforming  homogeneously under (\ref{gl54}) such as
\begin{equation}
\label{gl60}
D_{\mu} (B) a_{\nu}, \quad
F_{\mu\nu} (B) a_{\rho}, \dots .
\end{equation}
This implies that in all interesting cases $B_{\mu}$ enters only
inside covariant derivatives $D_{\mu}(B)$ or field strength tensors
$F_{\mu\nu}(B)$.

Consequently, the partial derivatives $\frac{\delta f(B,a)}{\delta B}$
and $\frac{\delta f(B,a)}{\delta a}$ also consist of structures
which are transformed
homogeneously.
As a result, if gauge variations
\begin{equation}
\label{gl61}
\frac{\delta^{(i)} f}{\delta \omega} =  \frac{\delta f}{\delta B}
\frac{\delta^{(i)} B}{\delta
\omega} +  \frac{\delta f}{\delta a}\frac{\delta^{(i)} a}{\delta
\omega}
\end{equation}
are to satisfy eq. (\ref{gl59}), the quantities $\frac{\delta^{(i)}B}{\delta
\omega}$ and $\frac{\delta^{(i)}a}{\delta \omega}$ must be also
constructed from  combinations transforming themselves  in accordance
with this equation
\footnote{It is not difficult to insure in all interesting
cases that indeed there is no "fine tuning" between the two terms of
eq.(\ref{gl61}) and each should satisfy the condition
separately.}.

With this in mind, we condclude that only the three first
variations (\ref{gl52})--(\ref{gl54}) lead to an invariant
generating functional $Z(J,B)$.
In all other cases,
$\frac{\delta^{(i)} a}{\delta \omega}$ and
$\frac{\delta^{(i)} B}{\delta \omega}$ consist of 'wrong' (in the
above sense) combinations of $B_{\mu}$-fields.

To illustrate  this point, we work out the standard
background gauge case
\begin{equation}
\label{gl62}
f^a (B,a) = (D_{\mu} (B) a_{\mu})^a = 0
\end{equation}
\begin{equation}
\label{gl63}
\frac{\delta f^a}{\delta B^b} = f^{acb} a^c =
i(T^c a^c)^{ab},\qquad\frac{\delta f^a}{\delta a^b}=D_{\mu}(B)^{ab}.
\end{equation}
Combining with eqs.(\ref{gl52})
\begin{equation}
\label{gl64}
\frac{\delta^{(1)} a_{\mu}^a}{\delta \omega^b} =
D_{\mu} (B+a)^{ab}, \quad \delta^{(1)} B_{\mu} = 0
\end{equation}
one gets the standard result \cite{1,2}
\begin{equation}
\label{gl65}
\left( \frac{\delta^{(1)} f}{\delta \omega} \right)^{ab} =
(D_{\mu} (B) D^{\mu} (B+a))^{ab}.
\end{equation}
In a similar way, eqs.(\ref{gl53}) leading to
\begin{equation}
\label{gl66}
\frac{\delta^{(2)} a_{\mu}^a}{\delta \omega^b} =
D_{\mu} (B)^{ab}, \quad
\frac{\delta^{(2)} B_{\mu}^a}{\delta \omega^b} =
-i (T^c a^c)^{ab}
\end{equation}
yield
\begin{equation}
\label{gl67}
\left( \frac{\delta^{(2)} f}{\delta \omega} \right)^{ab} =
(D_{\mu} (B) D_{\mu} (B) + a_{\mu} a_{\mu})^{ab},
\end{equation}
while eqs.(\ref{gl54})
\begin{equation}
\label{gl68}
\frac{\delta^{(3)} a_{\mu}^a}{\delta \omega^b} =
-i (T^c a^c)^{ab}, \quad
\frac{\delta^{(3)} B_{\mu}^a}{\delta \omega^b} =
D_{\mu} (B)^{ab}
\end{equation}
implies the final expression
\begin{equation}
\label{gl69}
\left( \frac{\delta^{(3)} f}{\delta \omega} \right)^{ab} =
-i [D_{\mu} (B), a_{\mu}]^{ab}.
\end{equation}
Simple but tedious calculations confirm our general conclusion that
these three forms of gauge  fixing of the $a_{\mu}$-
field do exhaust all possibilites to maintain the invariance
of $Z(J,B)$.

We observe that the determinants of the operators (\ref{gl65}) and
(\ref{gl67}) in the
leading order $(a_{\mu} = 0)$ are reduced to
\begin{equation}
\label{gl70}
det (-D^2_{\mu} (B)),
\end{equation}
which generalizes the abelian factor (\ref{gl26}) and represents two
propagating ghost polarizations interacting with the
background. Note that the determinant
of the function (\ref{gl69}) vanishes formally for $a_{\mu} = 0$.
In the following, we will always use the first kind of gauge
splitting (\ref{gl52}).

In Landau gauge, eq. (\ref{gl62}), one obtains
in leading order
\begin{equation}
\label{landau}
 Z^{(2)}(0,B) = [\frac{det(-D^2_{\mu}(B))}{\widetilde{det}(-\tilde
P_{\mu\rho}(B)
 (\delta_{\rho\sigma} D^2_{\lambda}(B) -
2 \hat{F}_{\rho\sigma}(B)) \tilde P_{\sigma\nu} (B))}]^{1/2}
\end{equation}
where $\tilde P_{\mu\nu}= \delta_{\mu\nu}-D_\mu \frac{1}{D^2_\lambda}
D_\nu$ and the determinant in the denominator indicates the
formal integration over three components of $a_\mu$ selected
by condition (\ref{gl62})
(see the next section for a similar detailed derivation).
We immediately conclude that the conditions of eq. (\ref{gl15})
are not satisfied. Ghosts are propagating and in $D_{\mu\nu}$
all three retained polarizations are propagating. Obviously, this
gauge is not unitary.

{}From the above analysis it is clear that the same arguments can be
applied to noncovariant gauges to maintain the invariance
under (\ref{gl50}). We must only insure that gauge
fixing function $f(B,a)$ is constructed from elements with $B_{\mu}$
entering via the (noncovariant) combinations of long derivatives and
field strength tensors. This is exactly what we will exploit when
constructing the nonabelian background generalizations of
Coulomb and axial gauges.

\Section{Coulomb background gauge}

We first apply the above considerations to obtain the
most suitable unitary generalization of the Coulomb gauge for
the perturbation theory in a background field which satisfies
eq. (\ref{gl18}). We will also derive the corresponding
propagator $D^{col}_{\mu\nu}(B)$ which is required for
applications.

As in the abelian case, we start with the quadratic
approximation for $F^2_{\mu\nu} (B+a)$ to write the generating
functional $Z^{(2)}$ in the form

\begin{equation}
\label{gl71}
Z^{(2)}(J,B) = \int Da_{\mu} \det
\frac{\delta^{(1)} f (B^{\omega}, a^{\omega})}{\delta \omega}
 |_{\omega_0}\delta (f(B,a))
 \exp \left[ +
\frac{1}{2} \int d^4x \left( \frac{1}{g^2} a^{a}_{\mu} K^{ab}_{\mu\nu}
(B) a^b_{\nu} - 2J^a_{\mu}a^a_{\mu} \right) \right]
\end{equation}
where
\begin{equation}
\label{gl72}
K^{ab}_{\mu\nu} = [D^2_{\rho} (B) \delta_{\mu\nu} -
D_{\mu} (B) D_{\nu} (B) - 2 \hat{F}_{\mu\nu} (B)]^{ab}
\end{equation}
and

\beq
\label{gl73}
\hat F^{ab}_{\mu\nu}\equiv f^{abc}F^c_{\mu\nu}.
\eeq

We neglected in the expansion of $F_{\mu\nu}^2(B+a)$
the contribution of the term linear in
$a_{\mu}$

\beq
\label{gl74}
4a^c_{\nu}D^{ca}_{\mu}(B)F^a_{\mu\nu}(B)
\eeq
assuming that it merely renormalizes the parameters of the effective
action $F^2_{\mu\nu}(B)$ expressed through the proper collective
coordinates. This is realized, for instance,  in the $1+1$ gas of kinks and
antikinks \cite{13}. We stress that this assumption is
necessary for the existence of a selfconsistent separation (\ref{gl1}) of the
$A_{\mu}$ field into the two parts.

In order to obtain the gauge fixing function $f(B,a)$ leading to
conditions (\ref{gl18}) we must cancel the mixed terms in
$a_{\mu} K_{\mu\nu}a_{\nu}$,
\begin{eqnarray}
\label{gl75}
& & a_0 K_{0i} a_i = - a_0 D_0(D_i + 2 D_0^{-1} \hat{F}_{0i}) a_i \nonumber \\
& & a_i K_{i0} a_0 = - a_i (D_i + 2 \hat{F}_{i0}D^{-1}_0) D_0 a_0
\end{eqnarray}
which leads to the following generalization of the abelian
condition (\ref{gl17})
\begin{equation}
\label{gl76}
f^c(B,a) = (D_i(B) + 2D_0^{-1}(B) \hat{F}_{0i} (B))^{cd} a_i^d
\equiv (N_i(B) a_i)^c.
\end{equation}
The operator $N_i^{ab}(B)$ selects the two transverse 'physical'
polarizations.
This gauge fixing function yields the second requirement (\ref{gl15}).

We note that the gauge fixing function
$f(B,a)$ of eq.(\ref{gl76}) can not be obtained by a simple substitution
of $\partial_i$ by $D_i(B)$ in the abelian form (\ref{gl17}).

With this $f(B,a)$, it is straighforward to insert the
corresponding gauge unity into the generating functional; we obtain
\begin{eqnarray}
\label{gl77}
& & Z^{(2)}(J,B) = \int D a_{\mu} det (-N_i(B) D_i(B+a))
\delta(N_i(B)a_i) \times \nonumber \\
& & \times \exp \left[ +\frac{1}{2} \int d^4x \left\{
\frac{1}{g^2} \left( a_0 D^2_i(B) a_0 +
a_i K_{ik} (B) a_k \right) - 2J_{\mu}a_{\mu} \right\} \right],
\end{eqnarray}
where
\begin{equation}
\label{gl78}
K_{ik}^{ab} (B) = (\delta_{ik} D^2_{\mu}(B)
- 2\hat{F}_{ik}(B) - D_i(B) D_k(B))^{ab}.
\end{equation}
In order to eliminate $\delta(N_i a_i)$
we introduce, as in the abelian case, the projector
\begin{equation}
\label{gl79}
P_{ik} (B) = \left( \delta_{ik} - N_i(B) \frac{1}{N_l^2(B)} N_k(B) \right),
\; N_i P_{ik} = P_{ki} N_i = 0,
\end{equation}
singling out two physical propogating polarizations among the three
$a_i$ components.
With
\begin{equation}
\label{gl80}
a_i = P_{ik} a_k + (1-P)_{ik}a_k \equiv a^{\perp}_i +
a^{\parallel}_i
\end{equation}
and
\begin{equation}
\label{gl81}
\delta(N_i a_i) = (det(-N^2_i))^{-1/2} \delta(a^{\parallel}_i),
\end{equation}
one arrives at
\begin{eqnarray}
\label{gl82}
&&Z^{(2)}(J,B)=\int
Da_0Da^{\perp}_i\frac{det(-N_i(B)D_i(B+a))}{det^{1/2}(-N^2_i(B))}
\exp\left[\frac{1}{2}\int d^4x\left( \frac{1}{g^2} \{a_0
D^2_ia_0\right.\right.+\nonumber\\ &&+\left.\left.
a^{\perp}_i(K_{ik}+N_iN_k) a^{\perp}_k \}-2J_0 a_0 - 2J^{\perp}_i
a^{\perp}_i \right)\right],
\end{eqnarray}
where we substituted  $K_{ik}\to\tilde
K_{ik}=K_{ik}+N_iN_k$ for a later convenience (note that $N_i
a_i^{\perp} = 0$).

Also here, only the transverse current
$J^{\perp}_i=P_{ik}J_k$ enters and thus
$D_{ik}$ will have the sandwiched form

\begin{equation}
\label{gl83}
D_{ik}(B)=P_{im}(B)(\tilde{K} (B)^{-1})_{mn}P_{nk} (B),
\end{equation}
generalizing the abelian  expression (\ref{gl24}).
Consequently after integration over $a_0, a^{\perp}_i$ one
obtains the final representation for $Z^{(2)}(J,B)$
\begin{eqnarray}
\label{gl84}
& & Z^{(2)}(J,B)=\left\{ \frac{det(-N_i(B)D_i(B+a))}{\left[det(-N^2_i
(B))det(-D^2_i(B))\right]^{1/2}} \right\}
(\widetilde{det}(-P_{im}\tilde K_{mn}P_{nk}))^{-1/2} \times \nonumber \\
&&\times\exp\left[-\frac{g^2}{2}\int(J_iD_{ik}J_k+J_0
D_{00}J_0)d^4x\right]
\end{eqnarray}
with
\begin{equation}
\label{gl85}
D_{00}(x,y|B)= <x| \frac{1}{D_i^2 (B)} |y>, \quad
D_{0i} = D_{i0} = 0
\end{equation}
and
\begin{equation}
\label{gl86}
D_{ik} (x,y|B) = <x| P_{ij} (B) (\delta_{jl} D^2_{\mu} -
2 \hat{F}_{jl} - D_j D_l + N_j N_l)^{-1} P_{lk} (B) |y>.
\end{equation}

The two equations (\ref{gl85}),(\ref{gl86}) are the central
result of our paper.

Let us now give a short interpretation of these expressions.
We conclude from eq.(\ref{gl84}) that ghosts are present but do not propagate
in time (the factor in curly brakets of eq.(\ref{gl84}) is equal to
unity when $B_{\mu} = 0$). Indeed the ghost kinetic term $(N_iD_i)$
differs from the corresponding one ($D^2_i$) for the nonpropagating
$a_0$--component
only by the spin dependent part (the last term in eq.(\ref{gl76})). In
contrast to eqs.(\ref{gl70}),(\ref{landau}) ghost are not
propagating and enter
only the loop corrections to the instantaneous
Coulomb exchange given by $D_{00}$ at the tree level.  Consequently,
this is a unitary gauge and ghosts do not introduce independent
propagating degrees of freedom (described by $D_{ik}$) which is
sufficient for the desired economical formulation of the
Hamiltonian approach.

We point out that the expression (\ref{gl86}) for $D_{ik}$ gives the correct
form for both spin-independent and spin-dependent  interactions of
transverse gluons (decoupled due to conditions (\ref{gl85}) from the
instantaneous component) with the background.

If the conditions (\ref{gl18}) are not imposed, there is a variety of
unitary gauges generalizing the abelian Coulomb one (\ref{gl17}).
Indeed every gauge where $\tilde N_i(B)$ has the same limit
\beq
\label{gl87}
\tilde{N}_i(B)\to\partial_i\quad \mbox{for} \quad B_{\mu} \to 0
\eeq
like $N_i(B)$ and obeys the second condition of eq. (\ref{gl15})
will lead to nonpropagating ghosts. Because of the coupling between
the $a_0$ and the two retained $a_i$-components ($D_{0i} \neq 0$)
the choice of the nonpropagating component is not ambigous
in this case.
It is the simple form (\ref{gl85}) and (\ref{gl86}) of
$D_{\mu\nu}$ which renders the choice (\ref{gl76}) of $N_i(B)$ the preferable
one for further applications with the multichannel Hamiltonian.

\section{Axial background gauges}

Next, we derive $Z(J,B)$ in quadratic approximation when the gauge
condition is imposed in the form
\begin{equation}
\label{gl88}
(n_{\mu} a_{\mu})^a = 0.
\end{equation}
As before, the term linear in $a_{\mu}$ (see eq. (\ref{gl74})) is neglected
also here.

First we analyse whether ghosts are also absent in these gauges
if there is a background. For this purpose it is sufficient to
consider $Z^{(2)}(0,B)$ and look for a representation in the
form of eq.(\ref{gl16}).
It will be shown that only the light cone $(n^2=0)$ axial gauge
remains ghost free in the presence of an arbitrary background.
Again we will obtain the propagator and give a short interpretation.\\

\subsection*{A. The case $n^2 \not= 0$}

As in the abelian case, for  $n^2 \not= 0$ it is
sufficient to consider $n = (1, \vec{0})$ or
$n = (0,1,\vec{0}_{\perp})$.
Both elementary choices are similar, we
take only the temporal gauge
\begin{equation}
\label{gl89}
n = (1, \vec{0}),
\end{equation}
for which in Euclidean space
\begin{equation}
\label{gl90}
Z^{(2)}(0,B) = \int Da_i det^{1/2} (-D^2_0 (B))
\exp [+\frac{1}{2} \int d^4 x
a_i K_{ik}(B) a_k],
\end{equation}
with
\begin{equation}
\label{gl91}
K^{ab}_{ik}(B) = [(D^2_0(B)+D^2_l(B)) \delta_{ik} - D_i (B) D_k (B)
- 2 \hat{F}_{ik} (B)]^{ab}.
\end{equation}
We note that in this gauge the ghost factor $det^{1/2}(-D^2_0(B))$
does not depend on the $a_{\mu}$-fields.
Equation (\ref{gl90}) immediately leads to
\begin{equation}
\label{gl92}
Z^{(2)}(0,B) = \left[ \frac{det (-D^2_0(B))}{det (-K_{ik}(B))} \right]^{1/2}.
\end{equation}
Again we see that $Z^{(2)}(0,B)$ is different for different background
gauges because the gaussian approximation to
$Z(0,B)$ implicit in perturbation
theory is not invariant under a change of the background gauges. This
raises the question whether there is a 'best' gauge.

As in the abelian case, the absence of ghosts
is equivalent to the existence of a projector
\begin{eqnarray}
\label{gl93}
& & P^{ab}_{ik}(B) = \delta^{ab} \delta_{ik}-\left( M_i(B)\frac{1}{M^2_l
(B)} M_k (B) \right)^{ab}, \quad P^2 = P  \nonumber \\
& & a_i = M_i \frac{1}{M^2_l} (M_k a_k) + P_{ik} a_k
\end{eqnarray}
such that the following generalization of the first condition
in (\ref{gl15}) holds (the second one is obviously satisfied)
\begin{equation}
\label{gl94}
det (-K_{ik}) = det (-D^2_0) \widetilde{det} (-P_{im} K'_{mn} P_{nk}).
\end{equation}
In other words, ghosts are absent if there exists an operator
$M_i^{ab}(B)$ such that the result of integration over
the instantaneous
component defined as $b = \frac{1}{\sqrt{M^2_i}}$
$(M_k a_k)$ $(Da_i = DbD(Pa)_{\perp})$
cancels exactly the ghost factor. The last condition results in
\begin{equation}
\label{gl95}
M_i D^2_0 \delta_{ik} M_k = M_i K_{ik} M_k
\end{equation}
which means in turn that the 3-dimentional kinetic matrix $\tilde{K}_{ik} =
K_{ik} - \delta_{ik}D^2_0$
\begin{equation}
\label{gl96}
\tilde{K}_{ik} = D^2_l \delta_{ik} - D_i D_k - 2 \hat{F}_{ik}
\end{equation}
has a continuum of zero modes
$\left( M_i \frac{1}{\sqrt{M^2_l}} b \right)$. This finally
leads to the
constraint
\begin{equation}
\label{gl97}
det [- \tilde{K}_{ik} ] = 0
\end{equation}
reproducing the condition (\ref{gl95}).

To demonstrate that eq.(\ref{gl95}) can not be satisfied without an extra
constraint on $F_{\mu \nu} (B)$ let us consider first
 the simpler $2+1$ case
where
\begin{equation}
\label{gl98}
det [- \tilde{K}_{ik}] \sim det (-D^2_1 + (D_1 D_2 + \hat{F}_{21})
D^{-2}_2 (D_2 D_1 + \hat{F}_{12})).
\end{equation}
The operator in eq.(\ref{gl98}) can be represented as
\begin{equation}
\label{gl99}
(\hat{F}_{12} D^{-1}_2 D_1 - D_1 D^{-1}_2 \hat{F}_{12})
+ \hat{F}_{12} D^{-2}_2 \hat{F}_{12},
\end{equation}
and one concludes that it vanishes in general only if
\begin{equation}
\label{gl100}
[D_i, \hat{F}_{ik}] = 0, \quad i,k = 1,2.
\end{equation}
This is not surprising because $\tilde{K}_{ik}$ of eqs.( \ref{gl96}),
(\ref{gl98})
coincides with the  kinetic
operator (\ref{gl72}) for the $1+1$ case and therefore has gauge zero modes
$D_i \omega$  \cite{1,2} in the presence of a background which satisfies
the classical $1+1$
equations of motion (\ref{gl100}). In the $2+1$ case at hand,
any classical background satisfies
\begin{equation}
\label{gl101}
[D_i, \hat{F}_{ik}] = -[D_0, \hat{F}_{0k}]
\end{equation}
which is equivalent to eq. (\ref{gl100}) only if in addition
\begin{equation}
\label{gl102}
[D_0, \hat{F}_{0k}] = 0
\end{equation}
is imposed.

It is not difficult to show that in $3+1$ dimensions,
eqs. (\ref{gl100}) with
$i,k = 1,2,3$ are necessary to satisfy eq.(\ref{gl95}).
Therefore, for $n^2 \not=
0$ it is impossible to cancel the ghost factor exactly for
an arbitrary background field, even if it is only classical.
Thus we conclude that in general the background
axial gauge (\ref{gl88}) is not ghost-free when $n^2 \neq 0$.

Still, for all gauges  where $M_i^2 (B)$ satisfies
the second condition (\ref{gl15}) and
\begin{equation}
\label{gl103}
\frac{1}{\sqrt{M^2_k(B)}} M_i (B) \to \frac{1}{\sqrt{\partial^2_k}}
\partial_i , \quad B_{\mu} \to 0
\end{equation}
one gets
\begin{equation}
\label{gl104}
Z^{(2)}(0,B) = \left\{ \frac{det (-D^2_0) det(-M^2_i)}{det
(M_i K_{ik} M_k)} \right\}^{1/2} \widetilde{det}^{-1/2} [(PK'P)_{ik}]
\end{equation}
and the factor in curly brakets becomes unity for $B_{\mu} = 0$
when we recover the abelian form (\ref{gl26}). Consequently, the only
propagating
part of $Z^{(2)}(0,B)$ arises from the integration over $(P_{ik} a_k)$ and
ghosts lead merely to loop corrections to the nonpropagating contribution
of the exchange. As a result, all such gauges (\ref{gl103}) can be called
unitary.

The impossibility to cancel $det^{1/2}
(-D^2_0)$ corresponds to the ambiguity in the
definition of the projector $P_{ik}$
onto transverse propagating polarizations since all $M_i(B)$ of
eq.(\ref{gl103})
are suitable (if the second condition (\ref{gl15}) is not violated).
Consequently, one can not select unambigously the non-propagating
component from the rest.

\subsection*{B. Light cone gauge $n^2 = 0$}

In this gauge with
\begin{equation}
\label{gl105}
n_{-} = 1, \quad n_{+} = n^{\perp}_i = 0,
\end{equation}
the quadratic approximation to  $Z(J,B)$ is given by
\begin{eqnarray}
\label{gl106}
& & Z^{(2)} (J,B) = \int Da_{-} Da_i \{det (-D^2_{+} (B)) \}^{1/2}
\times \nonumber \\
& & \times exp \left[+ \frac{i}{2} \int d^4 x \left\{ \frac{1}{g^2} a_{\mu}
K_{\mu \nu}(B) a_{\nu} + 2(J_{+} a_{-} + J_i a_i) \right\} \right],
\end{eqnarray}
where
\begin{equation}
\label{gl107}
\ba{ll}
K_{++} = -D^2_{+}, \; K_{+i} = -D_{+} D_{i} - 2\hat{F}_{+i}, \;
K_{i+} = -D_{i} D_{+} - 2\hat{F}_{i+}, \\
K_{ik} = D^2_{\mu} \delta_{ik} - D_i D_k - 2\hat{F}_{ik}
\ea
\end{equation}
and $i,k$ stand for the two perpendicular components.

{}From these equations one obtains after integration over the $a_{-}$
component
\begin{eqnarray}
\label{gl108}
& & Z^{(2)}(J,B) = \int Da_i \exp \left[ + \frac{i}{2g^2} \int d^4 x
\{a_i \tilde{K}_{ik} a_k - 2 g^2 a_i (K_{i+} K_{++}^{-1} J_{+} -
J_i) \} \right]\times \nonumber \\
& & \times \exp\left[-\frac{ig^2}{2}\int d^4x J_{+}(K_{++})^{-1}J_{+}\right],
\end{eqnarray}
where
\begin{equation}
\label{gl109}
\tilde{K}^{ab}_{ik} = K^{ab}_{ik} - (K_{i+} (K_{++})^{-1} K_{+k})^{ab}.
\end{equation}
We see that the ghost factor in eq. (6.36) is
cancelled exactly  by the  integration over $a_{-}$  as
in the abelian case. The absence of ghosts in
the light cone background gauge
makes it the most convenient one (among the axial gauges) for the
Hamiltonian approach built up in terms of
physical polarizations alone.

Integrating out the $a_i$-fields, one obtains $Z^{(2)}(J,B)$ in the
representation of eq.(\ref{gl10})
\begin{equation}
\label{gl110}
Z^{(2)}(J,B) = \{ \det \tilde{K}_{ik})\}^{-1} \exp \left\{ -
\frac{ig^2}{2} \int d^4 x (L_i (J) \tilde{K}^{-1}_{ik} L_k(J) +
J_{+} (K_{++})^{-1} J_{+}) \right\}
\end{equation}
where
\begin{equation}
\label{gl111}
L_{i}(J) = J_i - K_{i+} (K_{++})^{-1} J_{+}
\end{equation}
The physical interpretation of eq.(\ref{gl110}) is similar to the abelian
case (\ref{gl36}). The determinant in the
preexponent reproduces the statistical sum
$Z(0,B)$ in term of
 two propagating polarizations. In the exponent itself, the first
term gives the part of the propagator $D^{prop}_{\mu \nu} (B)$
responsible for these two propagating polarizations. The second term
supplies us with the modified instantaneous exchange

\begin{equation}
\label{gl112}
D^{inst}_{\mu \nu} (B) = \frac{n_{\mu} n_{\nu}}{D^2_+ (B)}
\end{equation}
generalizing the abelian expression (\ref{gl38}).We note that the choice of
physical propagating polarizations is unambiguous.

Similarly to $D_{ik}(B)$ of eq.(\ref{gl86}) in the Coulomb background
gauge, the piece $D^{prop}_{\mu \nu} (B)$ given by
\begin{equation}
\label{new}
D^{prop}_{\mu \nu} (B)=-(g^{\perp}_{\mu\rho} - n_{\mu}(nKn)^{-1}
(nK)^{\perp}_{\rho}) \tilde K^{-1}_{\rho\sigma}(g^{\perp}_{\sigma\nu} -
(Kn)^{\perp}_{\sigma}
(nKn)^{-1}n_{\nu})
\end{equation}
provides, in particular, the
light cone representation for the spin-dependent interactions of physical
polarizations with the background. We note that it is
more complicated than
in the Coulomb case.

\Section{Applications of unitary gauges}

As pointed out in the introduction, the basic motivation of our
work is to design a dynamical scheme to investigate QCD bound
states including gluonic excitations at the scale of confinement
("constituent gluons"). In this
section we sketch how the propagators can be used to this aim.

In order to simplify the discussion and to separate the
confining dynamics we are interested in here from
the effects of chiral symmetry breaking for light quarks,
we will consider the case of spinless quarks only.
For heavy quarks the
spin-dependent interactions can be related \cite{12} via a cluster
expansion to the same irreducuble correlators of background fields
which constitute the averaged Wilson loop. For light quarks these
interactions are connected to chiral symmetry breaking and an
additional information on the confining configurations is required.
Work in this direction is now in progress.

The Feynman-Schwinger representation allows one to write the Greens
function of the $q \overline q$ state in a path integral form.
For spinless quarks in the quenched approximation we have \cite{14}

\beq
\label{gl113}
G= \int d s_1 D z_1 d s_2 D z_2 \exp [ -K_1 - K_2 ] < W(C)>_{B+a}
\eeq
where
${<W(C)>}_{B+a}$ is the Wilson loop operator averaged over all
gluonic fields $A_\mu = B_\mu +a_\mu$
\begin{eqnarray}
\label{gl114}
& & <W(C)>_{B+a}= \frac{1}{Z} \int DB Da \det \frac{\delta^{(1)} f
(B^{\omega},a^\omega)}{\delta \omega}|_{\omega_0} \delta (f(B,a)) \times
\nonumber \\
& & \times \exp \left[ -  \frac{1}{4g_0^2} \int F^2_{\mu \nu} (B+a) d^4 x
\right] P \exp \left[ i \int_C (B_{\mu} + a_{\mu}) d x_{\mu} \right]
\end{eqnarray}
and the integration over $DB_\mu$ implies a summation over
all relevant collective coordinates excluded from $Da_\mu$ in the
standard way (we omit for simplicity the corresponding orthogonality
conditions ensured via the Faddeev--Popov trick). The contour $C$ consists
of quark and antiquark trajectories $z_1, z_2$ and $K_1$ and $K_2$ are the
standard quark kinetic terms \cite{14} whose explict form will not be
required here.

Quantum fluctuations $a_\mu$ can be taken into account by expanding
the path ordered exponent \\
$P\exp[i\int_C(B_{\mu}+a_{\mu})dx_{\mu}]$ in powers of
$a_\mu$. The result can be represented as
\beq
\label{gl116}
<W(C)>_{B+a} =
<W(C)>_B+\sum^{\infty}_{n=1}i^n\int
\left<W^{(n)}(C;x(1),\ldots,x(n))\right>_Bdx(1)\ldots dx(n)
\eeq
where $<W^{(n)}>_B$ is a Wilson loop with n insertions of the
$a_\mu$ fields \cite{4}. For example, the quantity
\begin{eqnarray}
\label{gl117}
& & \int<W^{(2)}>_Bdx(1)dx(2) = \nonumber \\
& & = g^2\int dx_{\mu}(1)dx_{\nu}(2)
\left<\Phi^{\alpha\beta}_{C_1}(x(1),x(2)|B)
D^{\delta\alpha;\beta\gamma}_{\mu\nu}(x(1),x(2)|B)
\Phi^{\gamma\delta}_{C_2}(x(2),x(1)|B)\right>_B+\nonumber\\
&& +O(g^3)
\end{eqnarray}
(with $D^{\delta\alpha;\beta\gamma}_{\mu\nu} =
t^{\delta\alpha}_a t^{\beta\gamma}_b D^{ab}_{\mu\nu}$)
describes (apart from the $O(g^3)$ corrections) the one gluon
exchange in Fig. 1a. Here $\Phi_{C_i}$ denotes necessary parallel
transporters in the fundamental representation along the subpathes $C_i$
of the initial contour C
\beq
\label{gl118}
\Phi_{C_i}(x,y|B)=P\exp\left[i\int^y_x(B^a_{\mu}t^a)dx_{\mu}\right]
\eeq
while $D_{\mu\nu}$ is the propagator of the fluctuations in the
background.

We emphasize that the subscript $B$ in eqs. (\ref{gl116}), (\ref{gl117})
refers to retaining only the $B_\mu$-field in all relevant
path ordered exponents. The averaging procedure is still performed
with the action density $F^2_{\mu \nu} (B+a)$ as in eq. (\ref{gl114}).

To incorporate $<W^{(n)}(C)>_B$ consistently into the bound
state formalism, a multichannel Hamiltonian approach has been
proposed \cite{9}.
Within this approach, time ordered diagrams  as in eq.(\ref{gl116})
can be reproduced by iterations of the effective Hamiltonian.
The problem is thus reformulated as that of a relativistic
many body system with mixed channels.

Under general circumstances, a selfconsistent application of
the multichannel
Hamiltonian formalism to a quantum field  theory requires \cite{8} the light
cone frame (coordinates).  For this purpose, the covariant
generalization \cite{15} of the Coulomb gauge (\ref{gl17}) can be used where
the new propagator takes the form (\ref{gl85}), (\ref{gl86}) in the
co-moving Minkowski
coordinates, obtained by a Lorentz transformation
in accordance with the total velocity of the
system $V_{\mu}$
\beq
\label{gl119}
n^{(0)}_{\mu}=V_{\mu}=\gamma(1,V,0_{\bot}),\quad
n^{(1)}_{\mu}=\gamma(V,1,0_{\bot}),\quad n^{(3,4)}_{\mu}=(0,0,\vec
n_{\bot}).
\eeq

One can express $n^{(\alpha)}_{\mu}$
via the conventional light cone coordinates to recover the conceptually
important
suppression of pair creation from the vacuum.
In the limit $V \to 1$, the new $D_{00}$ part expressed in terms
of light cone coordinates corresponds to the
instantaneous Coulomb exchange boosted from the rest frame to the
infinite momentum one \cite{15}.
 Consequently, it makes
sense to work with $D_{\mu\nu}^{Col}(B)$ in the form of eqs.(\ref{gl85}),
(\ref{gl86}), which is, in addition, somewhat simpler than $D_{\mu\nu}(B)$ of
eqs.(\ref{gl112}),(\ref{new}) in the
light cone gauge. Keeping this in mind, we are led to the following
strategy \cite{9}.

Take bound states with light quarks where the long
distance dynamics clearly dominates. Then one can treat
in Euclidean space the
hard and soft parts of the $a_\mu$-field separately as fast
and slow subsystems. First one integrates the hard part with
four-momentum squared $p^2_\mu  \geq \frac{1}{T^2_g}$
where $T_g$ is the scale at which the area law of $<W(C)>_B$
starts and the effective string between gluons and quarks forms.

Apart from the inducing new effective vertices, this also
leads to a renormalization of the running coupling constant at
the scale $\frac{1}{T_g}$ where it freezes due to the confining
vacuum \cite{4}. With this effective action the soft part (with Euclidean
four-momentum squared $p^2_\mu  \leq \frac{1}{T^2_g}$)
of the $a_\mu$-field is
handled, together with the $B_\mu$-field, in the following way.

At the initial step one must take into account the contributions from
the first term in eq.(\ref{gl116}), $<W(C)>_B$, plus the sum over all
orders in the instantaneous exchange $D_{00}(x,y|B)\sim\delta(x_+-
y_+)$ from the rest. In the following, we will always replace the integration
over
the background fields $B_\mu$ by the sum over irreducible
correlators of the field strength $F_{\mu\nu}(B)$  with the
help of  the cluster expansion \cite{14}. At the hadronic scale,
this enables one to
use the expansion in $\frac{T_g}{<r>}$ where $T_g\sim0.2-0.4\;fm$
\cite{14} is the
decay length of the correlators and $<r>$ is the caracteristic
distance between constituents. In this way one retains \cite{14}
the first few local terms (such as the area  and perimeter law
contributions) for quantities like  the  Wilson loop averaged over
$B_{\mu}$--fields only
\beq
\label{gl115}
<W(C)>_B = exp( -\sigma S + \rho l +...).
\eeq
where $S,l, \sigma$ and $\rho$ are the area,
length, string tension and effective mass respectively.
We note that at the level of the vacuum average, the contribution
of the higher derivative structures (such as the curvature)
is relatively suppressed for large smooth countours.

 Denoting this part of $<W(C)>_{B+a}$ as $<W^{(q\bar
q)}(C)>_{B+a}$, one can introduce the effective $q \overline q$
Lagrangian defined formally by

\beq
\label{gl120}
\int^T_0 L(q\bar q)dt_+=-\ln\left[\int
ds_1 D z_1 d s_2 D z_2\exp[-K_1-K_2]<W^{(q\bar q)}(C)>_{B+a}\right]
\eeq
After
continuing to Minkowski space, the Legendre transformation gives us
the Hamiltonian $H(q \overline q)$ for the two valence quark Fock
sector $\Psi(q\bar q)$.

We emphasize that $L(q \overline q)$ contains on the light
cone both the frozen flux
tube contribution \cite{16} arising from the area law asymptotics
(\ref{gl115}) of $<W(C)>_{B}$ 
and also the soft part of the
instantaneous Coulomb
exchange determined by $D_{00}(B)$ which describes the
interactions with the vacuum fields. Diagramatically,
it corresponds to Fig.2.

At the next stage, one can calculate those time ordered
diagrams of $<W(C)>_{B+a}$
which are obtained by the iterations of the the two-coupled channel
Hamiltonian

\beq
\label {gl121}
H^{(2)} =
\left(
\ba{ll}
H(q\bar q) & V(q\bar q\to q\bar qg)\\
V(q\bar qg\to q\bar q) & H(q\bar qg)
\ea
\right)
\eeq
acting on a Fock state vector
\beq
\label{gl122}
\Psi=
\left(
\ba{ll}
\Psi(q\bar q)\\
\Psi(q\bar qg)
\ea
\right)
\eeq
which includes also the state with one valence gluon (see Fig.1b).
For the case $B_{\mu} = 0$ this approximation to $H$ has been
considered in ref. \cite{17}.

Generally,
$H^{(2)}$ must be evaluated according to the scheme which we
discuss at the example of the simplest subset of the $D_{ik}$
iterations in Fig. 3. Time intervals with the propagating gluons are
described by the Hamiltonian $H(q \overline q g)$ of the first
excitation of the string 'frozen' during the rest of the time.  It
can be determined already from the
simple diagram of Fig. 4a (where the valence
gluon propagates during the entire time) in the same way \cite{16}
as $H(q\bar q)$.  This diagonal ansatz, for an arbitrary number
of gluons has been suggested recently in \cite{5}. We note that in
order to calculate $H(q \overline q g)$ fully, one must take into
account all instantaneous exchanges between valence quarks and the gluon
(as in Fig. 4b).

 Without these corrections, the elements of $H(q
\overline q g)$ are given by the Legendre transformation of the
Lagrangian matrix $L_{ik}(q \overline q g)$
(continued to Minkowski space) which is formally given by

\beq
\label{gl123}
\int^T_0 L_{ik}(q \bar q g) dt_+= - \ln\left[ds_1Dz_1ds_2Dz_2\exp[-K_1-K_2]
<\Phi(x,y|B)D_{ik}(x,y|B)\Phi(y,x|B)>_B\right]
\eeq
corresponding to Fig. 4a.
Apart from the kinetic and the spin-dependent terms,
there are at large distances
two frozen strings (Fig. 1b) connecting the $a_\mu$-gluon with
the quark and antiquark, respectively.
They appear explicitely after using the Feynman-Schwinger
representation of $D_{ik}(B)$ which implies a path ordered
exponent in the adjoint representation along the gluon
trajectories over which the path integral is performed
\cite{4}. Cluster expansion techniques \cite{dubi}
allow to reformulate the averaging in eq. (\ref{gl123}) as an average
for two Wilson loops in the fundamental representation with
the closed contours $(C_1+C_g),(C_1-C_g)$  bounded by the corresponding
quark and gluon trajectories. As a result, the area law asymptotics
for these Wilson loops $<W(C_1 \pm C_g)>_B$ induces at large distances
a (frozen) string between the gluon and each quark.
The sandwiched form (eq. (\ref{gl83}))
insures that there are only two transverse propagating
polarizations. Note that eq. (\ref{gl123}) relates  $D_{ik}$ of
eq. (\ref{gl86}) to both spin-independent and spin-dependent
effective interactions of the valence gluon.

Non-diagonal elements $V(q \overline q g \rightarrow q \overline q)$
can be obtained uniquely from the requirement that the amplitude of
Fig. 1a is reproduced in the second iteration of $H^{(2)}$ (see
ref. \cite{9} for details).

For the case $B_{\mu}=0$, the Hamiltonian
$H^{(2)}$
would lead \cite{17} to the existence of a continuum part of the spectrum.
Here the inclusion of the confining background allows to invoke (on light
cone in particular) the confining QCD string \cite{16} and its excitations
which is still not elaborated in the standard Hamiltonian approach
\cite{8} (see \cite{18} for a discussion).

In principle, one can continue this procedure to include
step by step higher order Fock states which are relevant for a
problem at hands.  There are some indications \cite{9} that there
exists, at least for the low lying bound states a dynamical parameter
(in addition to the moderate value of coupling constant $\alpha_{st}$
frozen \cite{4} due to the confining  background) which justifies
this expansion.\\[1cm]

\noindent
\un{\large\bf Acknowledgments}\\

This paper was prepared mainly during a few visits of one of the
authors (A.D.) to the University of Zurich. He would like to thank
all the staff of the Institute of Theoretical Physics for warm
hospitality extended to him. This work is supported by
Schweizerischer Nationalfonds, by the Russian
Fundamental Research Foundation (A.D.), grant N 93--02--14937, and
by INTASS.

\newpage

\section*{Figure captions}

\noindent
\begin{tabular}{p{1.5cm}p{15.5cm}}
Fig.1a & One $a_{\mu}$--gluon exchange contribution to
$<W(C)>_{B+a}$.\\[3mm]

Fig.1b & Modes of the QCD string arising (on light cone)
in the amplitude of Fig.1a.\\[3mm]

Fig.2 & The part of $<W(C)>_{B+a}$ responsible for the dynamics in
the valence quark Fock sector.\\[3mm]

Fig.3 & The simplest subset of exchanges from  the part of
$<W(C)>_{B+a}$ relevant for the two Fock states approximation (7.9)
to the Hamiltonian.\\[3mm]

Fig.4a & The simplest amplitude for reconstruction of $H(q\bar
qg)$.\\[3mm]

Fig.4b & Amplitudes with the instantaneous exchanges leading to a
corrected $H(q\bar qg)$.
\end{tabular}


\newpage

\begin{thebibliography}{99}

\bibitem{1} G.'tHooft, Phys. Rev. {\bf D14} (1976) 3432.

\bibitem{2} A.Polyakov, Nucl. Phys. {\bf B120} (1977) 429, \\
"Gauge fields and strings" Harwood academic publishers, 1987.

\bibitem{3} G.'tHooft, Nucl. Phys. {\bf B190} (1981) 455.

\bibitem{4} Yu.Simonov, Preprint HD--THEP--93--16.

\bibitem{5} Yu.Simonov in Proceedings of Hadron'93,
p.2629, edited by T.Bressani et. al.

\bibitem{14} H.Dosch, Yu.Simonov, Phys. Lett. {\bf B205} (1988) 339, \\
Yu.Simonov, Yad. Fiz. {\bf54} (1991) 192.

\bibitem{6} B.S.DeWitt, Phys. Rev. {\bf162} (1967) 1195, 1239;\\
G.'tHooft, Nucl. Phys. {\bf B62} (1973) 444;\\
G.'tHooft in Acta Universitatis Wratislavensis no.38, 12th Winter
School of Theoretical Physics in Karpacz; Functional and probabilistic
methods in quantum field theory. v.I (1975).

\bibitem{7} I.Tamm, J. Phys. (U.S.S.R.) {\bf9} (1945) 449;\\
S.Dancoff, Phys. Rev. {\bf78} (1950) 382.

\bibitem{8} S.Brodsky and H.Pauli, in Recent Aspects of Quantum
Fields, H.Mitter et. al., Eds., Lecture Notes in Physics, v.396
(Springer--Verlag, Berlin, 1991); \\
K.Wilson et. al., Phys. Rev. {\bf D49} (1994) 6720.

\bibitem{9} A.Dubin, in preparation.

\bibitem{10} Yu.Dokshitzer, D.Dyakonov, S.Troyan, Phys. Rep. {\bf58}
(1980) 269.

\bibitem{11} L.Abbott, Nucl. Phys. {\bf B185} (1981) 189;\\
W.Dittrich et. al., Phys. Lett. {\bf B128} (1983) 321, Phys. Lett.
{\bf B144} (1984) 99.

\bibitem{12} Yu.Simonov, Nucl. Phys. {\bf B324} (1989) 67.

\bibitem{13} E.Bogomolny, Phys. Lett. {\bf91B}  (1980) 431.


\bibitem{15} A.Dubin, J.Tjon, in preparation.

\bibitem{17} M.Krautg\"artner, H.Pauli, F.W\"olz, Phys. Rev. {\bf
D45} (1992) 3755.

\bibitem{16} A.Dubin, A.Kaidalov, Yu.Simonov, Phys. Lett. {\bf B343}
(1995) 310.

\bibitem{dubi} A.Dubin, Yu. Kalashnikova, preprint ITEP-40-94,
hep-ph 9406332, Yad. Fiz (in press).

\bibitem{18} R.Perry, Lectures presented at Hadrons'94, Gramado,
April, 1994.




\end{thebibliography}
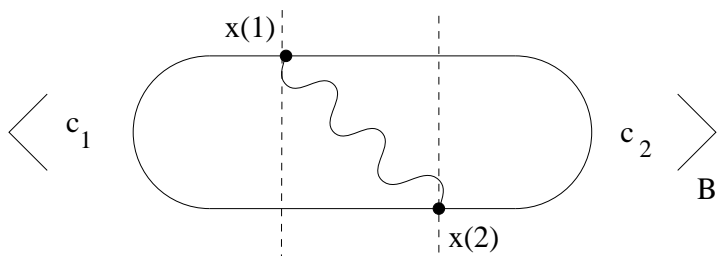
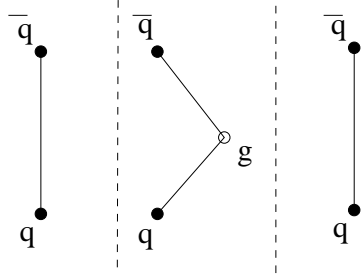
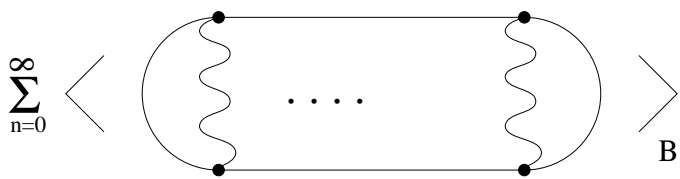
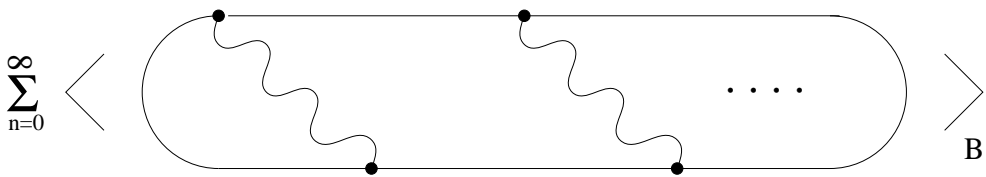
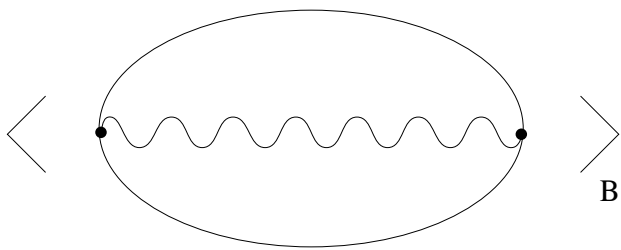
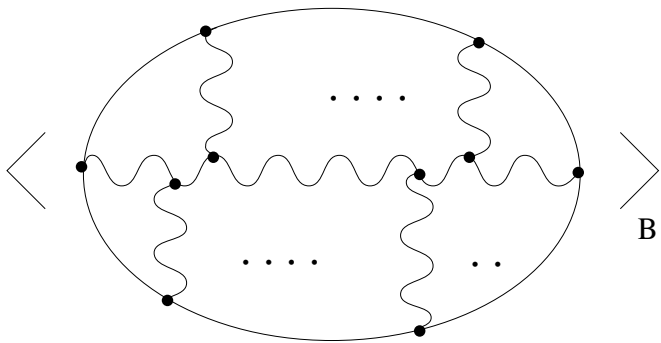
\end{document}